\documentclass[10pt]{article}
\usepackage{fullpage}

\usepackage{amsmath,amssymb}
\usepackage{tikz}
\usetikzlibrary{fit,automata,positioning,calc}
\usepackage{enumerate}
\usepackage{graphicx}

\begin{document}

\title{Successive Four-Dimensional Stokes-Space\\Direct~Detection\thanks{%
Submitted on Oct.~10, 2017 to the Optical Fiber Communication Conference and Exhibition, OFC 2018.  The authors are with the
Dept. of Electrical \& Computer Eng., Univ. of Toronto, 10
King's College Rd., Toronto, Ontario M5S 3G4, Canada.
Email: amir.tasbihi@mail.utoronto.ca; frank@ece.utoronto.ca}}
\author{Amir Tasbihi and Frank R. Kschischang}
\date{}

\maketitle

\begin{abstract}
We present a successive detection scheme for the fourth dimension in a
four-dimensional Stokes-space direct detection receiver.   At the
expense of a small number of electrical-domain computations, the
additional information rate can be substantial.
\end{abstract}

\section{Introduction}
\label{sec:intro}
Recently, the authors of~\cite{SVDD} have presented the Stokes-vector
direct detection transceiver as a promising candidate for noncoherent
short-haul fiber-optic communication. Their method was improved by
Morsy-Osman, et al.  in~\cite{4DSVDDPlant} to exploit all four
transmitted signal-space dimensions, but their proposed methods for
detecting the fourth dimension all suffer various practical drawbacks.
In this paper we propose a successive detection scheme with the same
optical receiver structure as~\cite{4DSVDDPlant}, but with some
additional computations in the electrical domain. We show that the
additional achievable information rate of the subchannel carried by the
fourth dimension can be quite substantial.

\section{Four-Dimensional Stokes-Space Direct Detection} 
\label{sec:4DSVDD}
Our method to detect three out of four used dimensions is the same as
in~\cite{4DSVDDPlant}, and is explained briefly here. For more details
the reader is referred to~\cite{4DSVDDPlant}. At time index $n$, the
transmitter sends two complex numbers $E_x[n]$ and $E_y[n]$, chosen from
a quadrature amplitude modulation constellation, over $X$ and $Y$
polarizations, respectively. The transmitted message is encoded in
$|E_x[n]|$, $|E_y[n]|$, $\arg\left(E_x[n]E_y^\ast[n]\right)$ and
$\arg\left(E_x[n]E_y^\ast[n-1]\right)$, where $z^\ast$ denotes the
complex conjugate of $z$. We refer to these respective quantities as the
first up to the fourth dimension.

\begin{figure}[htbp]
\centering
\definecolor{Col1}{rgb}{0.2,0.6,0.2}
\definecolor{Col2}{rgb}{0.4,0.2,0.4}
\definecolor{Col3}{rgb}{1,0,0}
\definecolor{Col4}{rgb}{0,0,1}
\definecolor{Col5}{rgb}{0,0.5,0.5}
\definecolor{Col6}{rgb}{0.45,0.35,0.2}
\definecolor{Col7}{rgb}{0.1,0.6,0.4}
\definecolor{Col8}{rgb}{0.5,0.5,0.}
\setlength{\tabcolsep}{2pt}
\begin{tabular}{ccc}
\begin{tikzpicture}[thick,scale=0.5, every node/.style={scale=0.5}]
\coordinate (A1) at (0,0);
\path (A1) +(1,0) coordinate (A4);
\path (A1) + (0,-1) coordinate (A2);
\path (A4) + (0,-1) coordinate (A3);
\draw (A1) to (A2);
\draw (A2) to (A3);
\draw (A3) to (A4);
\draw (A4) to (A1);
\draw (A1) to (A3);
\fill[color=Col1,fill=Col1,fill opacity=0.1](A1)--(A2)--(A3)--(A4);
\coordinate (B1) at ($(A1)!0.5!(A2)$);
\path (B1) + (-1,0) coordinate (B11);
\draw[->] (B11) to (B1);
\coordinate (B2) at ($(A1)!0.5!(A4)$);
\coordinate (B3) at ($(A2)!0.5!(A3)$);
\path (B2)+(0,2) coordinate (B22);
\path (B3)+(0,-2) coordinate (B33);
\coordinate (B3p) at ($(B3)!1/3!(B33)$);
\coordinate (B3pp) at ($(B3)!2/3!(B33)$);
\path (B3p)+(-0.6,0) coordinate (BB1);
\path (B3pp)+(-0.6,0) coordinate (BB2);
\path (B3pp)+(0.6,0) coordinate (BB3);
\path (B3p)+(0.6,0) coordinate (BB4);
\draw (BB1) to (BB2);
\draw (BB2) to (BB3);
\draw (BB3) to (BB4);
\draw (BB4) to (BB1);
\fill[color=Col6,fill=Col6,fill opacity=0.1](BB1)--(BB2)--(BB3)--(BB4);
\draw (B2) to (B22);
\draw (B3) to (B3p);
\draw (B3pp) to (B33);
\coordinate (D11) at ($(B2)!0.5!(B22)$);
\coordinate (D22) at ($(B3)!0.5!(B33)$);
\path (D11) + (1.5,0) coordinate (D1);
\path (D22) + (1.5,0) coordinate (D2);
\path (D1) + (1.5,0) coordinate (D4);
\path (D2) + (1.5,0) coordinate (D3);
\draw (D1) to (D2);
\draw (D2) to (D3);
\draw (D3) to (D4);
\draw (D4) to (D1);
\fill[color=Col2,fill=Col2,fill opacity=0.1](D1)--(D2)--(D3)--(D4);
\coordinate (D5) at ($(D1)!1/3!(D2)$);
\coordinate (D6) at ($(D1)!2/3!(D2)$);
\path (D5) +(-0.3,0) coordinate (D55);
\path (D6) +(-0.3,0) coordinate (D66);
\draw (D55) to (D5);
\draw (D66) to (D6);
\coordinate (D7) at ($(D3)!1/5!(D4)$);
\coordinate (D8) at ($(D3)!2/5!(D4)$);
\coordinate (D9) at ($(D3)!3/5!(D4)$);
\coordinate (D10) at ($(D3)!4/5!(D4)$);
\path (B22) + (6,0) coordinate (E22);
\path (B33) + (6,0) coordinate (E33);
\draw (B22) to (E22);
\path (B33)+(2,0) coordinate (L);
\path (L)+(0,0.4) coordinate (L1);
\path (L)+(0,-0.4) coordinate (L2);
\path (L1) +(1.2,0) coordinate (L4);
\path (L2) +(1.2,0) coordinate (L3);
\draw (L1) to (L2);
\draw (L2) to (L3);
\draw (L3) to (L4);
\draw (L4) to (L1);
\fill[color=Col5,fill=Col5,fill opacity=0.1](L1)--(L2)--(L3)--(L4);
\draw (B33) to (L);
\coordinate (L5) at ($(L3)!0.5!(L4)$);
\draw (L5) to (E33);
\coordinate (D555) at ($(B22)!(D55)!(E22)$);
\coordinate (D666) at ($(B33)!(D66)!(E33)$);
\draw (D555) to (D55);
\draw (D666) to (D66);
\path (D555)+(0,0.7) coordinate (D5555);
\path (D666)+(0,-0.7) coordinate (D6666);
\draw (D555) to (D5555);
\draw (D666) to (D6666);
\path (D7) +(0.7,0) coordinate (D77);
\path (D8) +(0.7,0) coordinate (D88);
\path (D9) +(0.7,0) coordinate (D99);
\path (D10) +(0.7,0) coordinate (D1010);
\draw (D7) to (D77);
\draw (D8) to (D88);
\draw (D9) to (D99);
\draw (D10) to (D1010);
\fill[color=Col3,fill=Col3,fill opacity=0.1](D77)--(D88)--(D99)--(D1010);
\path (D77)+(0,-0.25) coordinate (K1);
\path (D88)+(0,0.25) coordinate (K2);
\path (K1)+(1,0) coordinate (K4);
\path (K2)+(1,0) coordinate (K3);
\draw (K1) to (K2);
\draw (K2) to (K3);
\draw (K3) to (K4);
\draw (K4) to (K1);
\fill[color=Col3,fill=Col3,fill opacity=0.1](K1)--(K2)--(K3)--(K4);
\path (D99)+(0,-0.25) coordinate (KK1);
\path (D1010)+(0,0.25) coordinate (KK2);
\path (KK1)+(1,0) coordinate (KK4);
\path (KK2)+(1,0) coordinate (KK3);
\draw (KK1) to (KK2);
\draw (KK2) to (KK3);
\draw (KK3) to (KK4);
\draw (KK4) to (KK1);
\fill[color=Col3,fill=Col3,fill opacity=0.1](KK1)--(KK2)--(KK3)--(KK4);
\coordinate (K5) at ($(K3)!0.5!(K4)$);
\coordinate (K6) at ($(KK3)!0.5!(KK4)$);
\path (K5)+(0.5,0) coordinate (KK5);
\path (K6)+(0.5,0) coordinate (KK6);
\draw (K5) to (KK5);
\draw (K6) to (KK6);
\path (D5555)+(3,0) coordinate (E555);
\path (D6666)+(3,0) coordinate (E666);
\draw (D5555) to (E555);
\draw (D6666) to (E666);
\path (E555)+(0,0.3) coordinate (S1);
\path (E555)+(0,-0.3) coordinate (S2);
\path (S1) +(1,0) coordinate (S4);
\path (S2) +(1,0) coordinate (S3);
\draw (S1) to (S2);
\draw (S2) to (S3);
\draw (S3) to (S4);
\draw (S4) to (S1);
\fill[color=Col4,fill=Col4,fill opacity=0.1](S1)--(S2)--(S3)--(S4);
\coordinate (S5) at ($(S3)!0.5!(S4)$);
\path (S5)+(1,0) coordinate (S55);
\draw (S5) to (S55);
\path (E666)+(0,0.3) coordinate (SS1);
\path (E666)+(0,-0.3) coordinate (SS2);
\path (SS1) +(1,0) coordinate (SS4);
\path (SS2) +(1,0) coordinate (SS3);
\draw (SS1) to (SS2);
\draw (SS2) to (SS3);
\draw (SS3) to (SS4);
\draw (SS4) to (SS1);
\fill[color=Col4,fill=Col4,fill opacity=0.1](SS1)--(SS2)--(SS3)--(SS4);
\coordinate (SS5) at ($(SS3)!0.5!(SS4)$);
\path (SS5)+(1,0) coordinate (S66);
\draw (SS5) to (S66);
\path (D1) +(5,0) coordinate (M1);
\path (D2) +(5,0) coordinate (M2);
\path (M1) +(1.5,0) coordinate (M4);
\path (M2) + (1.5,0) coordinate (M3);
\draw (M1) to (M2);
\draw (M2) to (M3);
\draw (M3) to (M4);
\draw (M4) to (M1);
\fill[color=Col2,fill=Col2,fill opacity=0.1](M1)--(M2)--(M3)--(M4);
\coordinate (N1) at ($(M1)!1/3!(M2)$);
\coordinate (N2) at ($(M1)!2/3!(M2)$);
\path (N1) +(-3,0) coordinate (N111);
\path (N2) +(-3,0) coordinate (N222);
\coordinate (N11) at ($(N1)!(E22)!(N111)$);
\coordinate (N22) at ($(N2)!(E33)!(N222)$);
\draw (E22) to (N11);
\draw (E33) to (N22);
\draw (N11) to (N1);
\draw (N22) to (N2);
\coordinate (P1) at ($(M4)!1/5!(M3)$);
\coordinate (P2) at ($(M4)!2/5!(M3)$);
\coordinate (P3) at ($(M4)!3/5!(M3)$);
\coordinate (P4) at ($(M4)!4/5!(M3)$);
\path (P1) +(0.5,0) coordinate (P11);
\path (P2) +(0.5,0) coordinate (P22);
\path (P3) +(0.5,0) coordinate (P33);
\path (P4) +(0.5,0) coordinate (P44);
\draw (P1) to (P11);
\draw (P2) to (P22);
\draw (P3) to (P33);
\draw (P4) to (P44);
\path (P11) + (0,0.25) coordinate (Q1);
\path (P22) + (0,-0.25) coordinate (Q2);
\path (Q1) + (1,0) coordinate (Q4);
\path (Q2) + (1,0) coordinate (Q3);
\draw (Q1) to (Q2);
\draw (Q2) to (Q3);
\draw (Q3) to (Q4);
\draw (Q4) to (Q1);
\fill[color=Col3,fill=Col3,fill opacity=0.1](Q1)--(Q2)--(Q3)--(Q4);
\coordinate (Q5) at ($(Q3)!0.5!(Q4)$);
\path (Q5) +(0.5,0) coordinate (Q55);
\draw (Q5) to (Q55);
\path (P33) + (0,0.25) coordinate (Q11);
\path (P44) + (0,-0.25) coordinate (Q22);
\path (Q11) + (1,0) coordinate (Q44);
\path (Q22) + (1,0) coordinate (Q33);
\draw (Q11) to (Q22);
\draw (Q22) to (Q33);
\draw (Q33) to (Q44);
\draw (Q44) to (Q11);
\fill[color=Col3,fill=Col3,fill opacity=0.1](Q11)--(Q22)--(Q33)--(Q44);
\coordinate (Q6) at ($(Q33)!0.5!(Q44)$);
\path (Q6) +(0.5,0) coordinate (Q66);
\draw (Q6) to (Q66);
\node[circle,draw=black, fill=black, inner sep=0pt,minimum size=5pt] (b) at (D555) {};
\node[circle,draw=black, fill=black, inner sep=0pt,minimum size=5pt] (b) at (D666) {};
\draw (0,-0.6)node[anchor=north west]{\footnotesize PBS};
\draw (-0.3,1.2)node[anchor=north west]{\Large$F_{x}$};
\draw (-0.3,-2.35)node[anchor=north west]{\Large$F_{y}$};
\draw (-0.07,-1.75)node[anchor=north west]{\footnotesize $90^\circ$ rot};
\draw (2.55,-2.75)node[anchor=north west]{Delay};
\draw (2.4,0.3)node[anchor=north west]{$90^\circ$};
\draw (2.1,-0.2)node[anchor=north west]{Optical};
\draw (2.1,-0.7)node[anchor=north west]{Hybrid};
\draw (7.4,0.3)node[anchor=north west]{$90^\circ$};
\draw (7.1,-0.2)node[anchor=north west]{Optical};
\draw (7.1,-0.7)node[anchor=north west]{Hybrid};
\draw (4.2,0.33)node[anchor=north west]{BPD};
\draw (4.2,-0.85)node[anchor=north west]{BPD};
\draw (9,0.33)node[anchor=north west]{BPD};
\draw (9,-0.85)node[anchor=north west]{BPD};
\draw (4.85,2.95)node[anchor=north west]{PD};
\draw (4.85,-3.45)node[anchor=north west]{PD};
\draw (5.95,3.3)node[anchor=north west]{\LARGE$w_1$};
\draw (5.95,-3.7)node[anchor=north west]{\LARGE$w_2$};
\draw (5.2,0.7)node[anchor=north west]{\LARGE$w_3$};
\draw (5.2,-0.5)node[anchor=north west]{\LARGE$w_4$};
\draw (10.,0.7)node[anchor=north west]{\LARGE$w_5$};
\draw (10.,-0.5)node[anchor=north west]{\LARGE$w_6$};
\end{tikzpicture} & \begin{tikzpicture}[thick,scale=0.5, every node/.style={scale=0.5}]
\coordinate (A1) at (0,0);
\path (A1) +(1,0) coordinate (A4);
\path (A1) + (0,-1) coordinate (A2);
\path (A4) + (0,-1) coordinate (A3);
\draw (A1) to (A2);
\draw (A2) to (A3);
\draw (A3) to (A4);
\draw (A4) to (A1);
\draw (A1) to (A3);
\fill[color=Col1,fill=Col1,fill opacity=0.1](A1)--(A2)--(A3)--(A4);
\coordinate (B1) at ($(A1)!0.5!(A2)$);
\path (B1) + (-1,0) coordinate (B11);
\draw[->] (B11) to (B1);
\coordinate (B2) at ($(A1)!0.5!(A4)$);
\coordinate (B3) at ($(A2)!0.5!(A3)$);
\path (B2)+(0,2) coordinate (B22);
\path (B3)+(0,-2) coordinate (B33);
\coordinate (B3p) at ($(B3)!1/3!(B33)$);
\coordinate (B3pp) at ($(B3)!2/3!(B33)$);
\path (B3p)+(-0.6,0) coordinate (BB1);
\path (B3pp)+(-0.6,0) coordinate (BB2);
\path (B3pp)+(0.6,0) coordinate (BB3);
\path (B3p)+(0.6,0) coordinate (BB4);
\draw (BB1) to (BB2);
\draw (BB2) to (BB3);
\draw (BB3) to (BB4);
\draw (BB4) to (BB1);
\fill[color=Col6,fill=Col6,fill opacity=0.1](BB1)--(BB2)--(BB3)--(BB4);
\draw (B2) to (B22);
\draw (B3) to (B3p);
\draw (B3pp) to (B33);
\coordinate (D11) at ($(B2)!0.5!(B22)$);
\coordinate (D22) at ($(B3)!0.5!(B33)$);
\path (D11) + (1.5,0) coordinate (D1);
\path (D22) + (1.5,0) coordinate (D2);
\path (D1) + (1.5,0) coordinate (D4);
\path (D2) + (1.5,0) coordinate (D3);
\draw (D1) to (D2);
\draw (D2) to (D3);
\draw (D3) to (D4);
\draw (D4) to (D1);
\fill[color=Col2,fill=Col2,fill opacity=0.1](D1)--(D2)--(D3)--(D4);
\coordinate (D5) at ($(D1)!1/3!(D2)$);
\coordinate (D6) at ($(D1)!2/3!(D2)$);
\path (D5) +(-0.3,0) coordinate (D55);
\path (D6) +(-0.3,0) coordinate (D66);
\draw (D55) to (D5);
\draw (D66) to (D6);
\coordinate (D7) at ($(D3)!1/5!(D4)$);
\coordinate (D8) at ($(D3)!2/5!(D4)$);
\coordinate (D9) at ($(D3)!3/5!(D4)$);
\coordinate (D10) at ($(D3)!4/5!(D4)$);
\path (B22) + (6,0) coordinate (E22);
\path (B33) + (6,0) coordinate (E33);
\draw (B22) to (E22);
\path (B33)+(2,0) coordinate (L);
\path (L)+(0,0.4) coordinate (L1);
\path (L)+(0,-0.4) coordinate (L2);
\path (L1) +(1.2,0) coordinate (L4);
\path (L2) +(1.2,0) coordinate (L3);
\draw (L1) to (L2);
\draw (L2) to (L3);
\draw (L3) to (L4);
\draw (L4) to (L1);
\fill[color=Col5,fill=Col5,fill opacity=0.1](L1)--(L2)--(L3)--(L4);
\draw (B33) to (L);
\coordinate (L5) at ($(L3)!0.5!(L4)$);
\draw (L5) to (E33);
\coordinate (D555) at ($(B22)!(D55)!(E22)$);
\coordinate (D666) at ($(B33)!(D66)!(E33)$);
\draw (D555) to (D55);
\draw (D666) to (D66);
\path (D555)+(0,0.7) coordinate (D5555);
\path (D666)+(0,-0.7) coordinate (D6666);
\draw (D555) to (D5555);
\draw (D666) to (D6666);
\path (D7) +(0.3,0) coordinate (D77);
\path (D8) +(0.7,0) coordinate (D88);
\path (D9) +(0.3,0) coordinate (D99);
\path (D10) +(0.7,0) coordinate (D1010);
\draw (D7) to (D77);
\draw (D8) to (D88);
\draw (D9) to (D99);
\draw (D10) to (D1010);
\path (D88)+(0,-0.25) coordinate (K1);
\path (D88)+(0,0.25) coordinate (K2);
\path (K1)+(1,0) coordinate (K4);
\path (K2)+(1,0) coordinate (K3);
\draw (K1) to (K2);
\draw (K2) to (K3);
\draw (K3) to (K4);
\draw (K4) to (K1);
\fill[color=Col4,fill=Col4,fill opacity=0.1](K1)--(K2)--(K3)--(K4);
\path (D1010)+(0,-0.25) coordinate (KK1);
\path (D1010)+(0,0.25) coordinate (KK2);
\path (KK1)+(1,0) coordinate (KK4);
\path (KK2)+(1,0) coordinate (KK3);
\draw (KK1) to (KK2);
\draw (KK2) to (KK3);
\draw (KK3) to (KK4);
\draw (KK4) to (KK1);
\fill[color=Col4,fill=Col4,fill opacity=0.1](KK1)--(KK2)--(KK3)--(KK4);
\coordinate (K5) at ($(K3)!0.5!(K4)$);
\coordinate (K6) at ($(KK3)!0.5!(KK4)$);
\path (K5)+(0.5,0) coordinate (KK5);
\path (K6)+(0.5,0) coordinate (KK6);
\draw (K5) to (KK5);
\draw (K6) to (KK6);
\path (D5555)+(3,0) coordinate (E555);
\path (D6666)+(3,0) coordinate (E666);
\draw (D5555) to (E555);
\draw (D6666) to (E666);
\path (E555)+(0,0.3) coordinate (S1);
\path (E555)+(0,-0.3) coordinate (S2);
\path (S1) +(1,0) coordinate (S4);
\path (S2) +(1,0) coordinate (S3);
\draw (S1) to (S2);
\draw (S2) to (S3);
\draw (S3) to (S4);
\draw (S4) to (S1);
\fill[color=Col4,fill=Col4,fill opacity=0.1](S1)--(S2)--(S3)--(S4);
\coordinate (S5) at ($(S3)!0.5!(S4)$);
\path (S5)+(1,0) coordinate (S55);
\draw (S5) to (S55);
\path (E666)+(0,0.3) coordinate (SS1);
\path (E666)+(0,-0.3) coordinate (SS2);
\path (SS1) +(1,0) coordinate (SS4);
\path (SS2) +(1,0) coordinate (SS3);
\draw (SS1) to (SS2);
\draw (SS2) to (SS3);
\draw (SS3) to (SS4);
\draw (SS4) to (SS1);
\fill[color=Col4,fill=Col4,fill opacity=0.1](SS1)--(SS2)--(SS3)--(SS4);
\coordinate (SS5) at ($(SS3)!0.5!(SS4)$);
\path (SS5)+(1,0) coordinate (S66);
\draw (SS5) to (S66);
\path (D1) +(5,0) coordinate (M1);
\path (D2) +(5,0) coordinate (M2);
\path (M1) +(1.5,0) coordinate (M4);
\path (M2) + (1.5,0) coordinate (M3);
\draw (M1) to (M2);
\draw (M2) to (M3);
\draw (M3) to (M4);
\draw (M4) to (M1);
\fill[color=Col2,fill=Col2,fill opacity=0.1](M1)--(M2)--(M3)--(M4);
\coordinate (N1) at ($(M1)!1/3!(M2)$);
\coordinate (N2) at ($(M1)!2/3!(M2)$);
\path (N1) +(-3,0) coordinate (N111);
\path (N2) +(-3,0) coordinate (N222);
\coordinate (N11) at ($(N1)!(E22)!(N111)$);
\coordinate (N22) at ($(N2)!(E33)!(N222)$);
\draw (E22) to (N11);
\draw (E33) to (N22);
\draw (N11) to (N1);
\draw (N22) to (N2);
\coordinate (P1) at ($(M4)!1/5!(M3)$);
\coordinate (P2) at ($(M4)!2/5!(M3)$);
\coordinate (P3) at ($(M4)!3/5!(M3)$);
\coordinate (P4) at ($(M4)!4/5!(M3)$);
\path (P1) +(0.7,0) coordinate (P11);
\path (P2) +(0.3,0) coordinate (P22);
\path (P3) +(0.7,0) coordinate (P33);
\path (P4) +(0.3,0) coordinate (P44);
\draw (P1) to (P11);
\draw (P2) to (P22);
\draw (P3) to (P33);
\draw (P4) to (P44);
\path (P11) + (0,0.25) coordinate (Q1);
\path (P11) + (0,-0.25) coordinate (Q2);
\path (Q1) + (1,0) coordinate (Q4);
\path (Q2) + (1,0) coordinate (Q3);
\draw (Q1) to (Q2);
\draw (Q2) to (Q3);
\draw (Q3) to (Q4);
\draw (Q4) to (Q1);
\fill[color=Col4,fill=Col4,fill opacity=0.1](Q1)--(Q2)--(Q3)--(Q4);
\coordinate (Q5) at ($(Q3)!0.5!(Q4)$);
\path (Q5) +(0.5,0) coordinate (Q55);
\draw (Q5) to (Q55);
\path (P33) + (0,0.25) coordinate (Q11);
\path (P33) + (0,-0.25) coordinate (Q22);
\path (Q11) + (1,0) coordinate (Q44);
\path (Q22) + (1,0) coordinate (Q33);
\draw (Q11) to (Q22);
\draw (Q22) to (Q33);
\draw (Q33) to (Q44);
\draw (Q44) to (Q11);
\fill[color=Col4,fill=Col4,fill opacity=0.1](Q11)--(Q22)--(Q33)--(Q44);
\coordinate (Q6) at ($(Q33)!0.5!(Q44)$);
\path (Q6) +(0.5,0) coordinate (Q66);
\draw (Q6) to (Q66);
\node[circle,draw=black, fill=black, inner sep=0pt,minimum size=5pt] (b) at (D555) {};
\node[circle,draw=black, fill=black, inner sep=0pt,minimum size=5pt] (b) at (D666) {};
\draw (0,-0.6)node[anchor=north west]{\footnotesize PBS};
\draw (-0.3,1.2)node[anchor=north west]{\Large$F_{x}$};
\draw (-0.3,-2.35)node[anchor=north west]{\Large$F_{y}$};
\draw (-0.07,-1.75)node[anchor=north west]{\footnotesize $90^\circ$ rot};
\draw (2.55,-2.75)node[anchor=north west]{Delay};
\draw (2.4,0.3)node[anchor=north west]{$90^\circ$};
\draw (2.1,-0.2)node[anchor=north west]{Optical};
\draw (2.1,-0.7)node[anchor=north west]{Hybrid};
\draw (7.4,0.3)node[anchor=north west]{$90^\circ$};
\draw (7.1,-0.2)node[anchor=north west]{Optical};
\draw (7.1,-0.7)node[anchor=north west]{Hybrid};
\draw (4.35,0.63)node[anchor=north west]{PD};
\draw (4.35,-0.6)node[anchor=north west]{PD};
\draw (9.35,0.63)node[anchor=north west]{PD};
\draw (9.35,-0.6)node[anchor=north west]{PD};
\draw (4.85,2.95)node[anchor=north west]{PD};
\draw (4.85,-3.45)node[anchor=north west]{PD};
\draw (5.95,3.3)node[anchor=north west]{\LARGE$w_1$};
\draw (5.95,-3.7)node[anchor=north west]{\LARGE$w_2$};
\draw (5.2,1.3)node[anchor=north west]{\LARGE$w_3'$};
\draw (5.2,0)node[anchor=north west]{\LARGE$w_4'$};
\draw (10.2,1.3)node[anchor=north west]{\LARGE$w_5'$};
\draw (10.2,0)node[anchor=north west]{\LARGE$w_6'$};
\end{tikzpicture} & \scalebox{0.75}{\begin{tikzpicture}[>=stealth,circ/.style={circle,draw,fill=red!10!white,thick,minimum size=4em,inner sep=1pt}]
\node[circ] (x1) {$E_x[n]$};
\node[circ,left=8ex of x1] (x0) {$E_x[n-1]$};
\node[circ,below=8ex of x1] (y1) {$E_y[n]$};
\node[circ,left=8ex of y1] (y0) {$E_y[n-1]$};
\draw[thick,->] (y1) -- node[right] {$\theta[n]$} (x1);
\draw[thick,->] (y0) -- node[left] {$\theta[n-1]$} (x0);
\draw[thick,->] (x1) -- node[above,sloped] {$\eta[n]$} (y0);
\end{tikzpicture}} \\
(a) & (b) & (c) \end{tabular}
\caption{\footnotesize (a) The receiver structure of 4D-SSDD presented in~\cite{4DSVDDPlant};
PBS: polarization beam splitter, rot: rotator, BPD: balanced photo-detector,
PD: photo-diodes. (b) Second scheme uses four fewer PDs than the other scheme;
(c) Phase difference between simultaneous waveforms and between current and
previous waveforms.}
\label{fig:Receiver}
\end{figure}
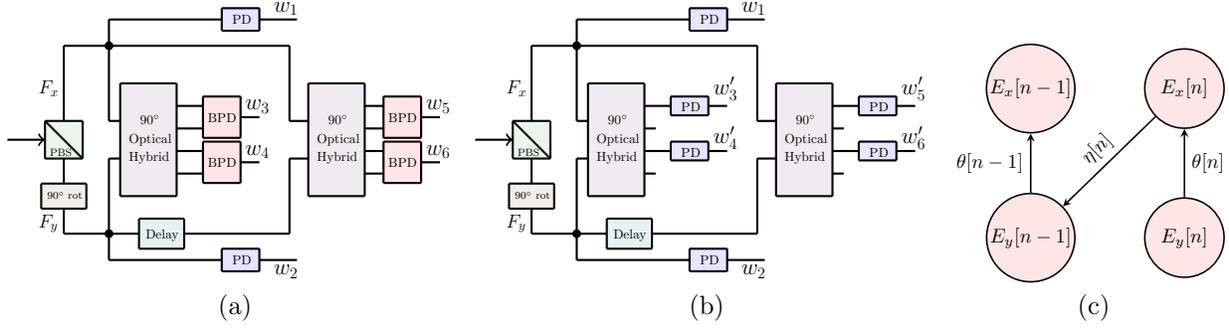

In single mode fibers, the output of the channel can be modeled
as~\cite{4DSVDDPlant} 
\[
\left[\begin{array}{c}
F_x[n]\\
F_y[n]
\end{array}\right]=\left[\begin{array}{cc}
a & b\\
-b^\ast & a^\ast
\end{array}\right]\left[\begin{array}{c}
E_x[n]\\
E_y[n]
\end{array}\right]+\left[\begin{array}{c}
z_x[n]\\
z_y[n]
\end{array}\right]=\left[\begin{array}{c}
K_x[n]\\
K_y[n]
\end{array}\right]+\left[\begin{array}{c}
z_x[n]\\
z_y[n]
\end{array}\right],
\]
where $F_x[n]$ and $F_y[n]$ are the  noisy received light in the two
polarizations, $z_x$ and $z_y$ are complex-valued circularly-symmetric
zero-mean additive white Gaussian noise
with variance $2\sigma^2$ due to the amplified spontaneous emission of the
preamplifier, $K_x[n]$ and $K_y[n]$ are the light components before
contaminating with the preamplifier noise, and the channel matrix is random
and unitary and is assumed to have a coherence time much larger than the
symbol duration. The received light is passed through the receiver shown in
Fig.~\ref{fig:Receiver}(a) to produce $w_1$ up to $w_6$. Without loss
of generality, we assume that the responsivity of the photo-diodes is unity,
so that
\begin{align*}
w_1[n] &=|F_x[n]|^2 , & w_2[n] & |F_y[n]|^2,\\  3[n] &= 2\Re(F_x[n]F_y^\ast[n]),   & w_4[n] &= 2\Im(F_x[n]F_y^\ast[n]),\\
w_5[n] &=2\Re(F_x[n]F_y^\ast[n-1]), & w_6[n]&= 2Im(F_x[n]F_y^\ast[n-1]),
\end{align*}
where $\Re(z)$ and $\Im(z)$ denote the real and imaginary parts of $z$. We
can estimate the channel matrix by transmitting three training symbols
$[1,0]^T,[1,1]^T$ and $[i,1]^T$, similar to the proposal in~\cite{SVDD}. 

We can recover the intensities of $E_{x}[n]$, $E_{y}[n]$ and their phase difference from the equation
\[
\left[\begin{array}{c}
w_1[n]\\
w_2[n]\\
w_3[n]\\
w_4[n]
\end{array}\right]=\left[\begin{array}{cccc}
|a|^2 & |b|^2 & \Re(ab^\ast) & -\Im(ab^\ast)\\
|b|^2 & |a|^2 & -\Re(ab^\ast) & \Im(ab^\ast)\\
-2\Re(ab) & 2\Re(ab) & \Re(a^2)-\Re(b^2) & -\Im(a^2)-\Im(b^2)\\
-2\Im(ab) & 2\Im(ab) & \Im(a^2)-\Im(b^2) & \Re(a^2)+\Re(b^2)
\end{array}\right]\left[\begin{array}{c}
|E_{x}[n]|^2\\
|E_{y}[n]|^2\\
2\Re(E_{x}[n]E_{y}^\ast[n])\\
2\Im(E_{x}[n]E_{y}^\ast[n])\\
\end{array}\right]+\left[\begin{array}{c}
\hat{z}_1\\
\hat{z}_2\\
\hat{z}_3\\
\hat{z}_4\\
\end{array}\right]
\]
where $\hat{z}_1$ up to $\hat{z}_4$ are correlated signal-dependent noise.
The maximum-likelihood detector would select $|E_x[n]|$, $|E_y[n]|$
and $\arg(E_x[n]E_y^\ast[n])$ to maximize
$f\left(w_1[n],w_2[n],w_3[n],w_4[n]\mid |E_x[n]|,|E_y[n]|,\arg\left(E_x[n]E_y^\ast[n]\right)\right)$, where $f(\cdot \mid \cdot)$ denotes the conditional probability density function (PDF).
As finding the exact conditional PDF is cumbersome, we approximate it with a
Gaussian with the same mean vector and covariance matrix. After estimation of
the channel matrix, $\left(|E_x[n]|,|E_y[n]|,\arg(E_x[n]E_y^\ast[n])\right)$
and $\left(|K_x[n]|,|K_y[n]|,\arg(K_x[n]K_y^\ast[n])\right)$ are in one-to-one 
correspondence, so we deal with the latter quantities.   If $\delta=\arg(K_x[n]K_y^\ast[n])$, the covariance matrix and
the mean vector are given as
\[
\left[
\setlength{\arraycolsep}{2pt}
\begin{array}{cccc}
4\sigma^4+4\sigma^2|K_x|^2 & 0 & 4\sigma^2|K_x||K_y|\cos(\delta) & 4\sigma^2|K_x||K_y|\sin(\delta)\\
0 & 4\sigma^4+4\sigma^2|K_y|^2 & 4\sigma^2|K_x||K_y|\cos(\delta) & 4\sigma^2|K_x||K_y|\sin(\delta)\\
4\sigma^2|K_x||K_y|\cos(\delta) & 4\sigma^2|K_x||K_y|\cos(\delta)& 8\sigma^4+4\sigma^2(|K_x|^2+|K_y|^2)& 0\\
4\sigma^2|K_x||K_y|\sin(\delta) & 4\sigma^2|K_x||K_y|\sin(\delta) & 0 & 8\sigma^4+4\sigma^2(|K_x|^2+|K_y|^2) \end{array}\right],~
\left[
\setlength{\arraycolsep}{0pt}
\begin{array}{@{}c@{}}
2\sigma^2+|K_x|^2 \\ 2\sigma^2+|K_y|^2 \\ 2|K_x||K_y|\cos(\delta)\\ 2|K_x||K_y|\sin(\delta)
\end{array} \right],
\]
respectively.  The symbol-error rate performance of this scheme is
shown in Section~\ref{sec:Results}.

Instead of the receiver shown in Fig.~\ref{fig:Receiver}(a), we can use the
one shown in Fig.~\ref{fig:Receiver}(b) which has a structure similar to
the Fig.~1b of ~\cite{SVDD}. As each balanced photo-detector has two
photo-diodes, the latter uses four fewer photo-diodes than the former. In this
case $w_3'[n]=w_1[n]+w_2[n]+\frac{1}{2}w_3[n]$ and we can recover $w_3$ in
the electrical domain, thereby reducing the number of
optical/electro-optical devices.
The same argument applies for $w_4', w_5'$ and $w_6'$. In this paper, the
calculations are based on $w_1$ up to $w_6$, which can be
obtained from the receiver shown in Fig.~\ref{fig:Receiver}(b) with
appropriate post-processing.

Until this point, we have detected the first three dimensions. To
detect the last dimension, $w_5$ and $w_6$ are used. These are related to
their
transmitted counterparts by
\begin{equation}
F_{x}[n]F_{y}^\ast[n-1]=\ell^{T}\left[\begin{array}{c}
E_{x}[n]E_{y}^\ast[n-1]\\
E_{y}[n]E_{x}^\ast[n-1]\\
E_{x}[n]E_{x}^\ast[n-1]\\
E_{y}[n]E_{y}^\ast[n-1]\\
\end{array}\right]+\hat{z}_{5}, \quad \ell=\left[\begin{array}{c}
a^2 \\
-b^2\\
-ab\\
ab
\end{array}\right],
\label{eq:FourthDimension}
\end{equation}
and where $\hat{z}_{5}$ is a signal dependent complex noise with correlated
components. As is apparent from (\ref{eq:FourthDimension}), the
received quantity is not just a function of
$E_{x}[n]E_{y}^\ast[n-1]$, but a function of three more beating terms.
In~\cite{4DSVDDPlant}, three suggestions for resolving this issue
are given:
(i) If $b=0$, then the channel has no rotation and in (\ref{eq:FourthDimension}), $F_{x}[n]F_{y}^\ast[n-1]$ is only a function of $E_{x}[n]E_{y}^\ast[n-1]$. This model is too simple and is not realistic.
(ii) In (\ref{eq:FourthDimension}) we have four unknowns, but just one
equation. So in order to solve a system of linear equations, we need three more
equations that can be obtained by measuring $F_{y}[n]F_{x}^\ast[n-1],
F_{x}[n]F_{x}^\ast[n-1]$ and $F_{y}[n]F_{y}^\ast[n-1]$ as well. By using the
same method as in Fig.~\ref{fig:Receiver}(a), it needs three more optical
hybrids and six more balanced photo-detectors which makes it quite complex. 
(iii) By feeding back the channel rotation matrix to the transmitter, we can pre-rotate the signal before transmission; however, this requires a feedback
channel.

\section{Successive Detection of the Fourth Dimension}
\label{sec:Results}

Here we present our method to detect the fourth dimension. If we examine
(\ref{eq:FourthDimension}) more carefully, we see that from just this
equation we can extract $E_{x}[n]E_{y}^\ast[n-1]$. If we denote the phase
difference of $E_{x}[n]$ and $E_{y}[n]$ by $\theta[n]$ and the phase difference
of $E_{x}[n]$ and $E_{y}[n-1]$ by $\eta[n]$, by using
Fig.~\ref{fig:Receiver}(c) we can re-write (\ref{eq:FourthDimension})
as
\begin{equation}
F_{x}[n]F_{y}^\ast[n-1]=\ell^T\left[\begin{array}{c}
|E_{x}[n]|\cdot|E_{y}[n-1]|\cdot \exp({i\eta[n]})\\
|E_{y}[n]|\cdot|E_{x}[n-1]|\cdot \exp({i(\theta[n]+\eta[n]+\theta[n-1])})\\
|E_{x}[n]|\cdot|E_{x}[n-1]|\cdot \exp({i(\eta[n]+\theta[n-1])})\\
|E_{y}[n]|\cdot|E_{y}[n-1]|\cdot \exp({i(\theta[n]+\eta[n])})\\
\end{array}\right]+\hat{z}_{5}.
\label{eq:MyderivedEquation}
\end{equation}
All the terms at time index $n-1$ are decoded and known. In addition, as
explained in Section~\ref{sec:4DSVDD}, we have detected the first three
dimensions and as a result $|E_{x}[n]|, |E_{y}[n]|$ and $\theta[n]$ also
are known. So the only unknown (besides noise) of the right-hand side of
(\ref{eq:MyderivedEquation}) is $\eta[n]$, which is the fourth dimension.
By rearranging the terms, it can be shown that
\[
\exp({i\eta[n]})=\frac{F_{x}[n]F_{y}^\ast[n-1]-\hat{z}_{5}}{\ell^Tv},
\quad v=\left[\begin{array}{c}
|E_{x,t}[n]|\cdot|E_{y,t}[n-1]|\\
|E_{y,t}[n]|\cdot|E_{x,t}[n-1]|\cdot \exp({i(\theta[n]+\theta[n-1])})\\
|E_{x,t}[n]|\cdot|E_{x,t}[n-1]|\cdot \exp({i\theta[n-1]})\\
|E_{y,t}[n]|\cdot|E_{y,t}[n-1]|\cdot \exp({i\theta[n]})\\
\end{array}\right]
\]
where $v$ is a known vector.

In summary, we perform successive detection.
First we decide on $|E_{x}[n]|,
|E_{y}[n]|$ and $\theta[n]$. Then we proceed to decide on $\eta[n]$ based on
$w_5, w_6$ and the decoded three dimensions by approximately maximizing
$f(w_5,w_6,|E_{x}[n]|,|E_{y}[n]|,\theta[n]\mid\eta[n])$ over all valid values
for $\eta[n]$, which is equivalent to maximizing
$f(w_5,w_6\mid\eta[n],|E_{x}[n]|,|E_{y}[n]|,\theta[n])$. Again, after
estimating the channel matrix and decoding $E_x[n], E_y[n]$ and $\theta[n]$,
there is a one-to-one correspondence between $\eta[n]$ and
$\arg(K_x[n]K_y^\ast[n-1])$. So if $\arg(K_x[n]K_y^\ast[n-1])=\alpha'$, then
the covariance matrix and mean vector for $w_5$ and $w_6$ are given,
respectively, as
$(8\sigma^4+4\sigma^2(|K_x[n]|^2+|K_y[n-1]|^2)) I_{2 \times 2}$
and
$[2|K_x[n]||K_y[n-1]|\cos(\alpha') , ~ 2|K_x[n]||K_y[n-1]|\sin(\alpha') ]^T$, where $I_{2\times 2}$ is the identity matrix.

\begin{figure}[htbp]
\centering
\begin{tabular}{cc}
\includegraphics[scale=0.75]{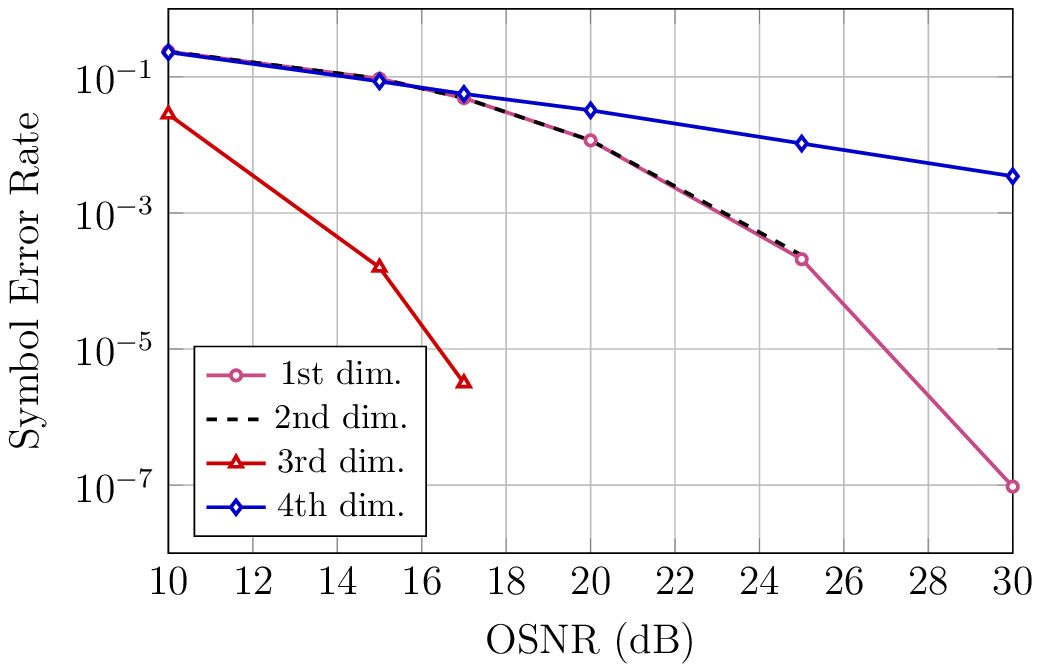} &
\includegraphics[scale=0.75]{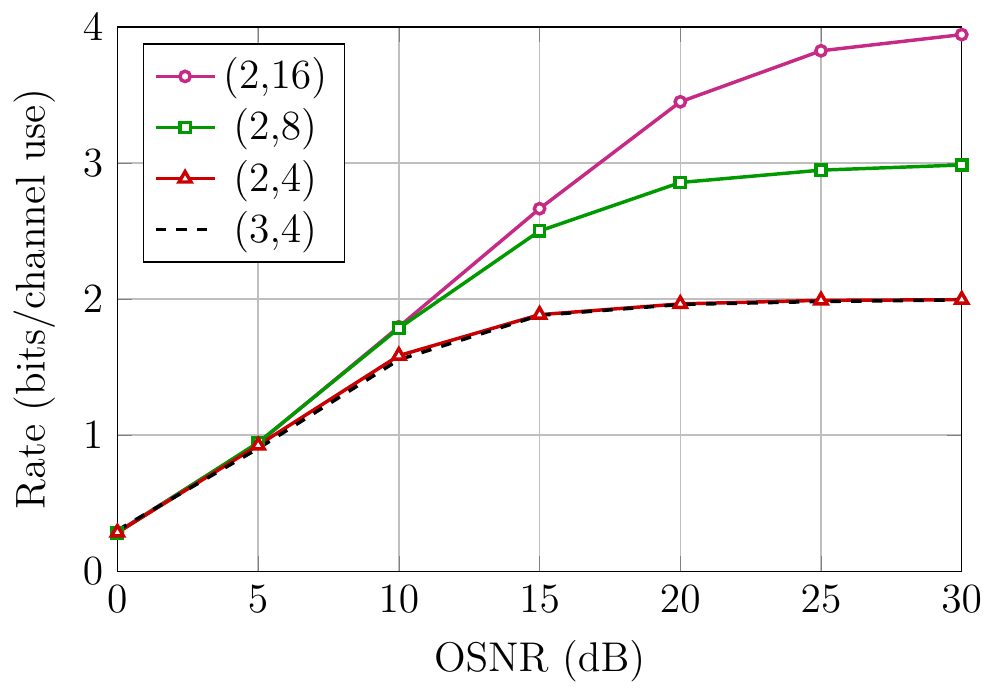}\\ (a) & (b) \end{tabular}
\caption{\footnotesize (a): SER of different dimensions for $2$-rings/$4$-ary PSK constellation by approximating the likelihood function as a Gaussian function. (b): Achievable rate of the fourth dimension for ($N_r,N_p$) constellations ($N_r$-ring/$N_p$-ary PSK).}
\label{fig:SERrates}
\end{figure}

We use an $N_r$-ring/$N_p$-ary PSK modulation, with equally-spaced
squared radii, as in~\cite{4DSVDDPlant}. The fiber loss is compensated by a
preamplifier
right before the PBS at the receiver (see Fig.~\ref{fig:Receiver}(a)).
The amplifier is the main source of noise.
The noisy waveform then enters the
receiver, and at the output we perform successive detection, using
a Gaussian approximation of the conditional PDF as previously
described.
The resulting symbol error rate is shown in
Fig.~\ref{fig:SERrates}(a).

In Fig.~\ref{fig:SERrates}(b) the average achievable rate of the
fourth dimension for several constellations is depicted. To obtain this figure,
we estimate mutual information by a discretization of the complex plane.
Fig.~\ref{fig:SERrates}(b) shows
that the achievable rate of the fourth
dimension can be substantial. For example, in the $2$-ring/$4$-ary PSK scheme
at an OSNR near 20~dB, the
achievable rate of the first three dimensions is $<4$ bits/channel-use,
while
the fourth dimension nearly provides an additional $2$ bits/channel-use.

\end{document}